\documentclass[aps,prl,twocolumn,superscriptaddress]{revtex4}
\usepackage{graphicx}
\usepackage{amsmath}
\begin{document}

\title{Strain-induced ferroelectricity in CaTiO$_3$ from first principles}

\author{C.-J. Eklund}

\affiliation{Department of Physics and Astronomy, Rutgers University, Piscataway, NJ 08854-8019, USA}

\author{C. J. Fennie} 

\affiliation{School of Applied and Engineering Physics, Cornell University, Ithaca, NY 14853-1501, USA}

\author{K. M. Rabe}

\affiliation{Department of Physics and Astronomy, Rutgers University, Piscataway, NJ 08854-8019, USA}

\date{\today}

\begin{abstract}
First principles calculations are used to investigate the effects of epitaxial strain on the structure of the perovskite oxide CaTiO$_3$, with particular focus on the stabilization of a ferroelectric phase related to a polar instability hidden in the orthorhombic equilibrium bulk $Pbnm$ structure but found in previous first-principles studies of the ideal cubic perovskite high-symmetry reference structure. At 1.5\% strain, we find an epitaxial orientation transition between the $ab$-$ePbnm$ phase, favored for compressive strains, and the $c$-$ePbnm$ phase. For larger tensile strains, a polar instability develops in the $c$-$ePbnm$ phase and an epitaxial-strain-induced ferroelectric phase is obtained with polarization along a $<$110$>$ direction with respect to the primitive perovskite lattice vectors of the square substrate. 
\end{abstract}






\maketitle

With the recent dramatic advances in the synthesis of coherent epitaxial films of complex oxides \cite{Schlomrev}, it is possible to maintain extremely high strains in thin film materials, often as much as 2--3\%. In some materials, such high strains can drive the system through a structural phase boundary to a novel phase with structure and properties distinct from those of the bulk equilibrium phase. First principles calculations can provide quantitative predictions about these novel phases and phase boundaries; in particular, such phases can be identified from examination of the lattice instabilities of the high-symmetry reference structure of the bulk phase \cite{RabeGA}, and their structure and properties predicted.

A prototypical example is strain-induced ferroelectricity in SrTiO$_3$. Starting from the bulk equilibrium paraelectric phase, strain-polarization coupling results in a polar instability beyond critical values of both tensile and compressive (001) epitaxial strain. This behavior was predicted by Landau theory \cite{Pertsev} and further analyzed in first principles investigations \cite{Antons,Eklund}. Experimental observation of epitaxial-strain induced ferroelectricity in SrTiO$_3$~\cite{STO,Eklund} demonstrates this strain-induced ferroelectricity both for compressive and tensile strain.

CaTiO$_3$ presents a greater challenge for the observation of strain-induced ferroelectricity. A relatively large temperature-dependent dielectric response has led to the characterization of CaTiO$_3$, like SrTiO$_3$, as an incipient ferroelectric \cite{Lemanov}. However, unlike SrTiO$_3$, CaTiO$_3$ has a strong tendency to oxygen-octahedron rotations, which tend to suppress polar lattice instabilities~\cite{Zhong,Sai}. Indeed, the bulk $Pbnm$ structure~\cite{Parlinski}, a structure type that includes a large number of other perovskite oxides, is obtained as a distortion of the high-symmetry cubic perovskite reference structure by freezing in components of the M$_3^+$ and R$_4^+$ oxygen octahedron modes (notation from Ref.~\onlinecite{Stokes}) involving rotation around [001] and tilting around [110], respectively, with additional changes in the lattice constants $a$, $b$ and $c$ and Ca displacements preserving the space group symmetry. All other known phases are also nonpolar: these include a paraelectric cubic structure at high temperatures and two intermediate phases (one tetragonal and one orthorhombic) \cite{Zelezny}. Less is known about high-pressure phases, though a transition to a nonpolar orthorhombic $Cmcm$ structure and, at higher pressures, to a nonpolar post-perovskite structure like that of MgSiO$_3$ have been proposed based on first-principles results \cite{Wupressure}. 
 
First-principles calculations of the full phonon dispersion relation of the cubic perovskite structure of CaTiO$_3$ show that in addition to the expected instability of M$_3^+$ and R$_4^+$, the cubic perovskite structure does have a strongly unstable polar $\Gamma_4^-$ mode~\cite{KSV,Parlinski,Cockayne}. Freezing in of this polar mode, which involves displacements of the Ca and Ti ions relative to the oxygen octahedron network, would yield a ferroelectric phase with nonzero spontaneous polarization. The fact that it does not contribute to the observed bulk phases suggests that it is inhibited by the oxygen octahedron rotations. If the rotations are artificially suppressed, as is possible in a first-principles calculation, the $\Gamma_4^-$ mode dominates and the resulting ferroelectric $P4mm$ phase is found to have a very large polarization~\cite{KSV,Nakhmanson}.

To stabilize a ferroelectric phase of CaTiO$_3$ under conditions realizable in the laboratory, it is necessary to change the balance of the competition between the octahedral rotations and the polar instability in favor of the latter. As in SrTiO$_3$, the polar instability in CaTiO$_3$ could be strengthened through the well-established sensitivity of the polar mode to strain in the titanates~\cite{Cohen}, specifically by tuning epitaxial strain~\cite{Serge, Ferro2006}. However, the oxygen-octahedron distortions inevitable in CaTiO$_3$ will couple to the epitaxial strain and polar instability, leading to a richer set of possibilities than in SrTiO$_3$. 

In this paper, we study the use of epitaxial strain to stabilize a ferroelectric phase of CaTiO$_3$. We focus on the orthorhombic bulk ground state, and investigate whether epitaxial strain can induce a polar instability, analogous to the behavior of SrTiO$_3$. The relatively low symmetry of the $Pbnm$ structure requires careful attention in imposing the epitaxial constraints, and introduces new features into the strain-induced ferroelectric state.


We performed density functional theory total-energy calculations within the LDA approximation using the VASP code ~\cite{version,vasp1,vasp2} with its supplied PAW potentials~\cite{paw1,paw2}. The plane-wave energy cutoff was 680 eV and the k-point grid was $8\times 8\times 6$. 

For the study of the effects of epitaxial strain, we carried out ``strained bulk" calculations, in which total-energy calculations are performed for the periodic crystal with appropriate epitaxial constraints imposed on the lattice parameters. In some cases, these are constraints which cannot be automatically imposed within the available VASP relaxation algorithms. In such cases, we developed an elastic energy expansion around the lowest energy $Pbnm$ structure that satisfies the epitaxial constraint by fitting to the energies of structures with small changes in strain (the latter structures not necessarily satisfying the epitaxial constraints). This elastic energy is then minimized with respect to strain subject to the epitaxial strain constraint; the resulting lattice parameters are then fixed in a total-energy calculation in which the internal structural parameters are relaxed. All structures were relaxed until the forces on the atoms were less than $2.5$ meV/\AA.

For selected structures, we compute the stability against zone-center modes by performing frozen phonon calculations in which single atoms are displaced by approximately $0.01$ \AA. From finite differences of the resulting forces, the force constant matrices are determined and subsequently diagonalized to obtain eigenfrequencies and eigenvectors.


Our results for the structure of the bulk orthorhombic ground state are given in Table \ref{wyckoff}. The structure has an energy 410 meV/f.u.\ lower than the ideal cubic perovskite structure. Consistent with previous first-principles calculations \cite{Cockayne,Parlinski}, we find good agreement between the computed structure and experiment, taking into account that in the local density approximation lattice constants typically tend to be underestimated by about one percent. 

\begin{table}  
\caption{\label{wyckoff} 
The Wyckoff parameters of the $Pbnm$ ground state and the $c$-$ePbnm$ 
structure at 3\% tensile strain.} 
\begin{ruledtabular} 
\begin{tabular}{rccc} 
& $Pbnm$ & $c$-$ePbnm$ \\ 
\hline 
Ca (4c)  &   $x=0.510, y=-0.047$       &  $x=0.509,  y=-0.042$ \\ 
Ti (4a)  &        $-$                  &         $-$           \\ 
O  (4c)  &   $x=0.081,  y=0.021$       &  $x=0.085,  y= 0.022$ \\ 
O  (8d)  &   $\begin{array}{c} x=0.207,y=0.292, \\ z=-0.043 \end{array}$ 
         &   $\begin{array}{c} x=0.218,y=0.283, \\ z=-0.044 \end{array}$ 
\end{tabular} 
\end{ruledtabular} 
\end{table} 

\begin{figure}
\includegraphics*[scale=0.3]{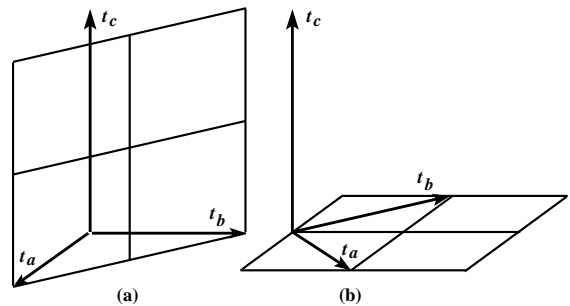}
\caption{\label{ePbnm} The two distinct relative orientations of the lattice vectors and the primitive perovskite substrate matching planes in the $Pbnm$ structure are shown for (a) the $ab$-$ePbnm$ phase and (b) the $c$-$ePbnm$ phase.}
\end{figure}

Next, we investigate the effects of epitaxial strain on the $Pbnm$ phase. As in Ref. \onlinecite{Zayak}, we designate the strained phases as $ePbnm$, where the prefix $e$ denotes ``epitaxial.'' We consider epitaxial strain on a square lattice substrate, corresponding to a (001) perovskite surface. To allow direct comparison with experiment despite the lattice constant underestimate discussed above, we define epitaxial strain relative to $a_0$ = 3.77 \AA, which is the cube root of the computed volume per formula unit of the relaxed $Pbnm$ structure. In the $Pbnm$ structure, there are two symmetry-inequivalent primitive perovskite (001) planes, as shown in Fig. \ref{ePbnm}. Thus, there are two distinct orientations for an epitaxial film: the first, with $\boldsymbol{t}_c$ in the matching plane and $\boldsymbol{t}_a$ and $\boldsymbol{t}_b$ out of the plane ($ab$-$ePbnm$, Figure \ref{ePbnm} (a)), and the second, with $\boldsymbol{t}_c$ normal to the matching plane ($c$-$ePbnm$, Fig. \ref{ePbnm}(b)).

We compute the total energies for these two orientations for epitaxial strains ranging from $-$1.5\% to 4\% \cite{footnote-strain}. For $c$-$ePbnm$, the $c$ lattice parameter and internal structural parameters are relaxed at each strain, maintaining the $Pbnm$ symmetry; the $a$ and $b$ lattice parameters are fixed by the constraint. The epitaxial constraint allows for $\boldsymbol{t}_c$ not to be normal to the matching $ab$-plane and tilting $\boldsymbol{t}_c$ could lower the energy. However, an elastic analysis for the $-$1.5\% and 4\% cases showed that tilting $\boldsymbol{t}_c$ does not lower the energy and we assumed this to be true for the intermediate strains as well. $ab$-$ePbnm$ has lower symmetry than $c$-$ePbnm$; that is, distinguishing one of the two (110) planes removes space group symmetries, resulting in a space group $P2_1/m$; the constraint fixes $|\boldsymbol{t}_c|$ and $|\boldsymbol{t}_b-\boldsymbol{t}_a|$, as well as the condition $\boldsymbol{t}_c\cdot (\boldsymbol{t}_b-\boldsymbol{t}_a)$ = 0. To optimize the lattice parameters for this case, we use the elastic energy expansion method described in the previous section.

\begin{figure}
\includegraphics*[scale=0.35]{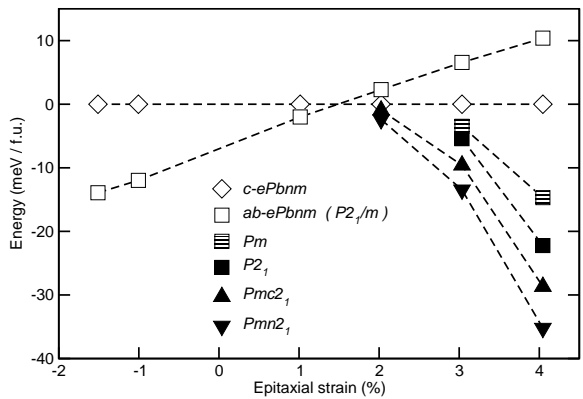}
\caption{\label{totalenergy} Total energy per five-atom formula unit for various epitaxially constrained structures as a function of misfit strain. At each strain, the energy of the $c$-$ePbnm$ structure is taken as the zero of energy. The connecting lines are a guide to the eye.}
\end{figure}

The results are shown in Fig. \ref{totalenergy}. $ab$-$ePbnm$ is favorable for compressive strains and $c$-$ePbnm$ is favored with increasing tensile strains. Within this subspace of nonpolar structures, there is an epitaxial orientation transition at 1.5\%.

Next, we turn to the stability of the $ePbnm$ phases against symmetry-breaking distortions, with special attention to polar phonons. Previous computation of the phonon frequencies for the bulk equilibrium structure showed three low-frequency polar phonons at 94 cm$^{-1}$, 88 cm$^{-1}$ and 89 cm$^{-1}$, with induced polarizations along $a$, $b$ and $c$, respectively~\cite{Cockayne}. These phonons are expected to be sensitive to changes in strain, based on known polarization-strain coupling in calcium titanate~\cite{Dieguez}. 

We first consider $c$-$Pbnm$ with 4\% tensile strain, with computed structural parameters reported in Table 1. A zone-center frozen phonon computation for this structure showed four unstable phonons, the lowest two, at 213i~cm$^{-1}$ and 209i~cm$^{-1}$, being polar and generating structures with space groups $Pmn2_1$ (polarization along $a$) and $Pmc2_1$ (polarization along $b$), respectively. In both cases, the orientation of polarization with respect to the primitive perovskite axes is along the $<$110$>$ directions. The energies of the structures for these two space groups, optimized under the epitaxial strain constraint, are 35 meV/f.u.~and 28 meV/f.u.~below $c$-$Pbnm$, respectively, with polarizations 0.46 C/m$^2$ and 0.45 C/m$^2$ computed using Born effective charges and atomic displacements~\cite{Born}. Thus, at 4\% tensile strain, we predict strain-induced ferroelectricity in CaTiO$_3$.

For the full range of strains, the unconstrained internal structural parameters were optimized within these two polar space groups. Only in the 4\% case was $|\boldsymbol{t}_c|$ re-optimized for the polar structures; this only had a marginal effect on the energy and the polarization compared to when the relaxed value for the $c$-$ePbnm$ structure was used, presumably because the polarizations are in the $ab$-plane. At compressive strain, the nonpolar $c$-$ePbnm$ structure is stable against polar distortions, and no ferroelectricity is observed. For tensile strain, the ferroelectric instabilities first appear at 2\% strain, and the energy gain and polarization of the optimized ferroelectric phases grow with increasing strain.

Let us now consider $ab$-$ePbnm$ with 4\% tensile strain. We looked for polar instabilities in this structure by displacing atoms in such a way that the resulting, lowered symmetry allowed for nonzero polarization, and then relaxing the internal structural parameters while keeping the lattice parameters fixed at their $ab$-$ePbnm$ values. This procedure revealed two structures with polarizations in the matching plane of 0.33 C/m$^2$ along $c$ (space group $P2_1$) and 0.40 C/m$^2$ along the $ab$ diagonal (space group $Pm$), respectively. (The polarizations were obtained using Born effective charges and atomic positions~\cite{Born}.) The former is the lowest in energy but still above $Pmn2_1$ and $Pmc2_1$, see Fig. \ref{totalenergy}. This procedure was also carried out for epitaxial strains of 3\% and $-$1.5\%. No polar phase was found in the latter case.


The mechanism of strain-induced ferroelectricity in the $Pbnm$ phase is closely related to that for SrTiO$_3$, which similarly has a high strain sensitivity of the low-frequency polar mode. However, the effect in SrTiO$_3$ is equally strong for compressive as for tensile strain, as elongation of the unit cell produced by compressive strain destabilizes the polar mode as effectively as elongation in the in-plane direction for tensile strain. This is a direct consequence of the fact that SrTiO$_3$ is cubic in the paraelectric phase; the cubic-tetragonal transition being at low temperatures and the rotational distortion not being strong enough to inhibit the strain-enhanced polar instability. 

In CaTiO$_3$, in contrast, the rotational instabilities are much stronger and the resulting distortions are much larger. No ferroelectric $ePbnm$ phase was found for compressive strain, despite elongation of the unit cell along the direction normal to the surface; this presumably is due to inhibition by the pattern of octahedral rotations. This highlights the idea that in a nonpolar low-symmetry phase, unlike in a cubic phase, the relationship between the crystal axes and the epitaxial constraints is very important, different choices yielding quite distinct structures and coupling to potential instabilities. 

To investigate the relative importance of enhancing the polar instability compared to suppressing the rotational instabilities, we analyzed the structural parameters of the $c$-$ePbnm$ phase as a function of epitaxial strain. The amplitudes of the M$_3^+$ and R$_4^+$ modes change surprisingly little in the range of strains reported here, suggesting that the dominant mechanism of the strain-induced ferroelectricity is the strain enhancement of the polar instability. 

Epitaxial strain in the $Pbnm$ is not the only possible avenue to a ferroelectric phase. Another promising approach is to replace the rotational pattern in the $Pbnm$ structure with a different pattern which allows the gain associated with the octahedral rotation instability but which is less inhibitory to the polar modes; this will be discussed in a separate publication \cite{future}.

In comparing these predictions with experiments on epitaxial strained CaTiO$_3$, it is important to keep in mind that our approach considers only the effect of strain on the ground state structure and properties. In a real thin film, especially the ultrathin films needed to sustain very high strains, other factors can affect the observed phase, including temperature, the atomic arrangements at the substrate and the film-substrate interface, relaxation, reconstruction, and adsorption at the free surface, and defects and impurities in the film itself. However, the tendency to ferroelectricity with increasing tensile strain is clear in our results, and to the extent that these other factors do not act dominantly against it, we expect ferroelectricity in CaTiO$_3$ to be observed at experimentally accessible strains. Investigations are currently in progress \cite{SchlomCTO}.


In summary, we have performed first-principles calculations for epitaxially strained structures of CaTiO$_3$. At 1.5\% strain, we find an epitaxial orientation transition between the $ab$-$ePbnm$ phase, favored for compressive strains, and the $c$-$ePbnm$ phase, favored for tensile strains. For sufficiently large tensile strains, a polar instability, which was hidden in the equilibrium bulk structure, develops in both phases. The epitaxial-strain-induced ferroelectric phase lowest in energy originates in the $c$-$ePbnm$ phase and has the polarization along a $<$110$>$ direction with respect to the primitive perovskite lattice vectors of the square substrate.

\section{Acknowledgments}

We would like to thank M.\ H.\ Cohen, V.\ Gopalan, D.\ R.\ Hamann, D.\ G.\ Schlom, and D.\ Vanderbilt for useful discussions. This work was supported by NSF MRSEC DMR-0820404 and ONR N00014-09-1-0302. K.\ M.\ R. would also like to thank the Aspen Center for Physics, where part of this work was carried out.


\begin{thebibliography}{99}
\bibitem{version} Versions 4.6.28  and 4.6.26 of VASP were used.
\end{thebibliography}
\end{document}